\documentclass[11pt]{article}

\usepackage{algorithmic,algorithm}
\usepackage{color,multicol,graphics,epsfig,epstopdf}
\usepackage{amsmath}
\usepackage{amsthm}
\usepackage{amssymb}
\usepackage{float}
\usepackage{graphicx,url}
\usepackage{cite}
\usepackage{wasysym}
\usepackage{fullpage}

\usepackage{color}
\definecolor{Red}{rgb}{1,0,0}
\definecolor{Blue}{rgb}{0,0,1}
\def\red{}
\def\blue{}

\def\mm{\mathbf{m}}
\def\pp{\mathbf{p}}
\def\qq{\mathbf{q}}
\def\zz{\mathbf{z}}
\def\00{\mathbf{0}}

\def\aa{\mathbf{a}}

\def\bb{\mathbf{b}}
\def\approxmm{\mathbf{\tilde{m}}}
\def\approxm{\tilde{m}}

\def\approxp{\tilde{p}}

\def\prob#1#2{\mbox{$\mathbf P\mathbf r$}_{#1}\left[ #2 \right]}
\def\nnz{\mbox{\rm nnz}}
\def\sparse{\mbox{\rm Sparse}}
\def\rangeIndicator{\mbox{\rm rangeIndicator}}
\def\Reals#1{\mathbb{R}^{#1}}

\begin{document}

\title{Multi-Scale Matrix Sampling and Sublinear-Time PageRank
  Computation\thanks{Accepted to \textit{Internet Mathematics} journal for publication. An extended abstract of this paper appeared in
     {\em WAW 2012} (pages 41-53) under
     the title ``A Sublinear Time Algorithm for PageRank Computations''.}}

\author{Christian Borgs\thanks{Microsoft Research New England. Email:
    borgs@microsoft.com.} \and  Michael Brautbar\thanks{ Computer and
    Information Science Dept., University of Pennsylvania. Email:
    brautbar@cis.upenn.edu.} \and  Jennifer Chayes\thanks{Microsoft
    Research New England. Email: jchayes@microsoft.com.} \and
  Shang-Hua Teng\thanks{Computer Science Dept., University of Southern
    California. Email: shanghua@usc.edu.
    {\blue Supported in part by NSF grants CCF-1111270 and CCF-0964481.}}
}

\date{}
\maketitle



\newtheorem{theorem}{Theorem}[section]
\newtheorem{corollary}[theorem]{Corollary}
\newtheorem{lemma}[theorem]{Lemma}
\newtheorem{claim}[theorem]{Claim}
\newtheorem{fact}[theorem]{Fact}
\newtheorem{proposition}[theorem]{Proposition}
\newtheorem{conjecture}{Conjecture}
\newtheorem{property}{Property}
\newtheorem{observation}[theorem]{Observation}
\newtheorem{remark}{Remark}
\newtheorem{discussion}{Discussion}
\newtheorem{definition}{Definition}



\newcommand{\ignore}[1]{}
 \newcommand{\todo}[1]{\begin{center}\fbox{\parbox[t]{0.95\textwidth}{#1}}\end{center}}
\newcommand{\mb}[1]{{[\textbf{MB: #1}]}}
\def\11{\mathbf{1}}
\newcommand{\graph}           {$G = (V,E)$}
\newcommand{\othergraph}      {\ensuremath{{\cal H}} }
\newcommand{\util}[1]         {\text{\it utility}(#1)}
\newcommand{\idealized}[2]    {{\text{\it ideal}}_{#1}(#2)}
\newcommand{\totaldeg}[1]     {\text{\it deg}(#1)}
\newcommand{\indeg}[1]        {\text{\it in-deg}(#1)}
\newcommand{\outdeg}[1]       {\text{\it out-deg}(#1)}
\newcommand{\inset}[1]        {\ensuremath{\text{\it I}_{#1}}}
\newcommand{\outset}[1]       {\ensuremath{\text{\it O}_{#1}}}
\newcommand{\totaltriang}[1]  {\ensuremath{\Delta(#1)}}
\newcommand{\intriang}[1]     {\ensuremath{\Delta_{I}(#1)}}
\newcommand{\outtriang}[1]    {\ensuremath{\Delta_{O}(#1)}}
\newcommand{\inouttriang}[1]  {\ensuremath{\Delta_{I,O}(#1)}}
\newcommand{\CC}[1]           {\text{\it CC}(#1)}
\newcommand{\distsum}[1]      {\text{\it SumOfShortestPaths}(#1)}
\newcommand{\far}[1]          {\text{\it far}(#1)}

\maketitle
\setcounter{page}{0}
\begin{abstract}
{\blue A fundamental  problem arising in many applications in Web science
and social network analysis is the problem of identifying all nodes in a
network whose PageRank exceeds a given threshold $\Delta$.  In this
paper, we study the probabilistic version of the problem where given an
arbitrary approximation factor $c>1$, we are asked to output a set $S$
of nodes such that with high probability, $S$ contains all nodes of
PageRank at least $\Delta$, and no node of PageRank smaller than
$\Delta/c$.} {\blue We call this problem {\sc SignificantPageRanks}.

We develop a nearly optimal, local algorithm
  for the problem with runtime complexity $\tilde{O}(n/\Delta)$
  on networks with $n$ nodes, where the tilde hides a
  polylogarithmic factor.
We show that any algorithm for solving this problem
  must have runtime of ${\Omega}(n/\Delta)$,
  rendering our algorithm
  optimal up to logarithmic factors.
Our algorithm has sublinear time complexity
  for applications including Web crawling and Web search
  that require efficient identification of nodes whose PageRanks are
  above a threshold $\Delta = n^{\delta}$, for some constant $0 <\delta <
  1$.}

Our algorithm comes with two main technical contributions.
The first is a multi-scale sampling scheme for a basic matrix
problem that could be of interest on its own.
{\blue For us, it appears as
an abstraction of a subproblem we need to tackle in
order to solve
the {\sc SignificantPageRanks} problem,
but we hope that this abstraction will be
  useful in designing fast algorithms for
  identifying nodes that are
  significant beyond PageRank measurements.

  In the abstract matrix problem
it is assumed that one can access an}
unknown {\em right-stochastic matrix}
  by querying its rows, where the cost of a query
  and the accuracy of the answers depend on a precision parameter
  $\epsilon$.
At a cost propositional to $1/\epsilon$,
  the query will return a list of $O(1/\epsilon)$ entries and their
  indices that provide an $\epsilon$-precision approximation of the
  row.
Our task is to find a set that contains all columns
  whose sum is at least $\Delta$, and omits any column whose sum is
  less than $\Delta/c$.
Our multi-scale sampling scheme solves this problem with
  cost $\tilde{O}(n/\Delta)$,
  {\blue while traditional sampling algorithms would take
  time $\Theta((n/\Delta)^2)$.}

Our second main technical contribution is a new local
  algorithm for approximating personalized PageRank,
  which is more
  robust than the earlier ones developed in
  \cite{JehW03,AndersenCL06} and is highly
  efficient particularly for networks
  with large in-degrees or out-degrees.

Together with our multiscale sampling scheme we are able to optimally solve the
 {\sc SignificantPageRanks} problem.\\

\end{abstract}
\newpage

\section{Introduction}

A basic problem in network analysis is to identify the set of network
 nodes that are ``significant.''
For example, they could be the significant Web pages that provide
   the authoritative contents in Web search;
they could be the critical proteins in a protein interaction network;
and they could be the set of people (in a social network)
  most effective to seed the
  influence for online advertising.
As the networks become larger, we need more efficient algorithms to
  identify these ``significant'' nodes.

\paragraph{Identifying Nodes with Significant PageRanks}

The meanings and measures of significant vertices depend on the
  semantics of the network and the applications.
In this paper, we focus on a particular measure of significance --- the
\textit{PageRank} of the vertices.

Formally, the PageRank (with restart constant,
  also known as the teleportation
 constant, $\alpha$) of a Web page is proportional to the
 probability that the page is visited by a
 random surfer who explores the Web using the following simple random
 walk:
 at each step, with probability $(1-\alpha)$ go to a random Web page
 linked to from the current page, and with probability $\alpha$,
 restart the process from a uniformly chosen Web page.
For the ease of presentation of our later results, we consider a normalization of PageRank so that
  the sum of the PageRank values
  over all vertices is equal to $n$, the number
  of vertices in the network,
\[ \sum_{u\in V} \text{PageRank}(u) = n.\]

PageRank has been used by the Google search engine and has found
 applications in a wide range of data analysis problems \cite{Berkhin05,BrinP98}.
In this paper, we consider the  following
 natural problem of finding vertices with ``significant'' PageRank:

\begin{quote}
{\bf {\sc SignificantPageRanks}:}
{\em Given a network $G =(V,E)$,
     a threshold value $1\leq \Delta \leq |V|$
     and a positive constant  $c>1$,
     compute a subset $S\subseteq V$ with thWe
     property that $S$ contains all vertices of
     PageRank at
     least $\Delta$ and
     no vertex with PageRank less than $\Delta/c$. }
\end{quote}

For the corresponding algorithmic problem we assume that the network
 topology is described in the \emph{sparse} representation of an
 (arbitrarily ordered) adjacency list for each vertex, as is natural
 for sparse graphs such as social and information networks.
We are
 interested in developing an efficient
local algorithm \cite{SpielmanTengLocalClustering,AndersenCL06,AndersenBCHMT08} for the problem
 in the context of Web applications.
The algorithm is only allowed to
  randomly sample out-links of previously accessed nodes in addition to
  sampling nodes uniformly at random from the network.
This model is highly suitable for
  PageRank maintenance in Web graphs and online information networks.

{\blue As the main contribution of this paper,
  we present a nearly optimal, local algorithm for
  {\sc SignificantPageRanks}.
}
The running time of our algorithm is $\tilde{O}({n}/{\Delta})$.
We also show that any
  algorithm for  {\sc SignificantPageRanks}
  must have query complexity as well
  as runtime complexity ${\Omega}({n}/{\Delta})$.
Thus, our algorithm is optimal up to a logarithmic factor.
{\blue
Note that when $\Delta = \Omega(n^\delta)$, for some constant
$0< \delta < 0$, our algorithm has sublinear time complexity.
}

Our {\sc SignificantPageRanks} algorithm applies a multiscale
  matrix sampling scheme that uses a fast
  Personalized PageRank estimator (see
  below) as its main subroutine.

\paragraph{Personalized PageRanks}

While the PageRank of a vertex captures the importance of
the vertex as collectively assigned by all vertices in the network,
one can use the distributions of the following random walks to define the pairwise
contributions of significance \cite{Haveliwala03}:
given a teleportation probability $\alpha$ and a starting vertex $u$
in a network $G=(V,E)$, at each step,
 with probability $(1-\alpha)$ go to a random neighboring vertex,
 and with probability $\alpha$, restarts the process from $u$.
For $v\in V$, the probability that $v$ is visited by this random
  process, denoted by $\text{PersonalizedPageRank}_{u}(v)$, is $u$'s personal PageRank contribution of significance to $v$.
It is not hard to verify that
\begin{eqnarray*}
& \mbox{$\forall u\in V$, \ } & \sum_{v\in V}
  \text{PersonalizedPageRank}_{u}(v)  = 1; \mbox{\ and }\\
& \mbox{$\forall  v\in V$, \ } & \text{PageRank}(v)=  \sum_{u\in V}
   \text{PersonalizedPageRank}_{u}(v).
\end{eqnarray*}

Personalized PageRanks has been widely used
 to describe personalized behavior of Web users \cite{PageBMW98}
 as well as for developing good network clustering techniques \cite{AndersenCL06}.
As a result, fast algorithms for computing or
  approximating personalized PageRank are quite useful.
One can approximate PageRanks and personalized PageRanks by
  the power method \cite{Berkhin05}, which involves costly matrix-vector
  multiplications for large scale networks.
Applying effective truncation,
  Jeh and Widom \cite{JehW03} and Andersen, Chung, and
  Lang \cite{AndersenCL06}
  developed personalized PageRank approximation algorithms
  that can find an $\epsilon$-additive approximation
  in time proportional to the product of $\epsilon^{-1}$ and
  the maximum in-degree in the graph.

\paragraph{Multi-Scale Matrix Sampling}\label{sec:intro:matrix}

Following the matrix view of the personalized PageRank
  formulation of Haveliwala \cite{Haveliwala03} and the subsequent
  approximation of algorithms \cite{JehW03,AndersenCL06},
  we introduce a matrix problem whose solution would lead to
  fast PageRank approximation and sublinear-time algorithms
  for {\sc SignificantPageRanks}.

In the basic form of this matrix problem, we consider a
  blackbox model for accessing an unknown $n\times n$ {\em right-stochastic matrix},
  in which we can only make a query of the following form:
  {\em matrixAccess}($i,\epsilon$), where $1 \leq i \leq n$ and $\epsilon \in (0,1]$.
This query will return, with high probability, a list of $O(1/\epsilon)$ entry-index pairs
  that provide an $\epsilon$-precise  approximation of row $i$ in the
  unknown matrix: For each $1 \leq j \leq n$, if $(p,j)$ is in the list of
entry-index pairs returned by
  {\em matrixAccess}($i,\epsilon$), then $|p-m_{i,j}|\leq \epsilon$,
  where $m_{i,j}$ is the $(i,j)^{th}$ entry of the unknown matrix; otherwise
  if there is no entry containing index $j$,
  then $m_{i,j}$ is guaranteed to be at most $\epsilon$.
Further, the cost of this query is propositional to $1/\epsilon$.
We will refer to this blackbox model as the {\em sparse and
  approximate row access model}, or SARA model for short.

We now define the basic form of our matrix problem:
\begin{quote}
{\bf {\sc SignificantMatrixColumns}:}
{\em Given an $n\times n$ {\em right-stochastic matrix} $M$ in the
  SARA model, a threshold $\Delta$ and a positive constant $c>1$,
  return a subset of columns $S\subseteq V$ with the
  property that $S$ contains all columns of sum at
  least $\Delta$ and no column with sum less than $\Delta/c$.}
\end{quote}

There is a straightforward connection between {\sc SignificantMatrixColumns}
 and {\sc SignificantPageRanks}.
Following \cite{AndersenBCHMT08}, we define a matrix PPR (short for
PersonalizedPageRank) to be the $n \times n$ matrix, whose $u^{th}$ row
is $$\text{PersonalizedPageRank}_{u}(\cdot).$$
Clearly PPR is a right-stochastic matrix and
  for $1 \leq v \leq n$, $\text{PageRank}(v)$
  is equal to the sum of the $v^{th}$
column in PPR.
Therefore, if we can solve the {\sc SignificantMatrixColumns} problem
  with cost $\tilde{O}(n/\Delta)$ and  also
  solve the problem of computing an $\epsilon$-additive approximation
  of personalized PageRank in $\tilde{O}(\log(n)/\epsilon)$
  time,
  then we are able to solve {\sc SignificantPageRanks} in
  $\tilde{O}(n/\Delta)$ time.

In this paper, we analyze a
 multi-scale sampling algorithm for {\sc SignificantMatrixColumns}.
The algorithm selects a set of
   precision parameters $\{\epsilon_1, ...,\epsilon_h\}$
   where $h$ grows linearly with  $n/\Delta$ and
  $\epsilon_i = i/h$.
It then makes use of the sparse-and-approximate-row-access queries
  to obtain approximations of randomly sampled rows.
For each $i$ in range $1\leq i \leq h$,  the algorithm makes
    $\tilde O(1)$ (depending on the desired success probability) row-access queries to get
    a good approximation to the contribution of column elements
    of value of order $\epsilon_i$.
We show that with probability $1-o(1)$, the multi-scale sampling scheme
  solve {\sc SignificantMatrixColumns}
   with cost $\tilde{O}(n/\Delta)$.

{\blue
While we could present our algorithm directly on PPR, we hope this matrix
 abstraction enables us to better highlight the two key algorithmic
  components in our fast PageRank approximation algorithm:
\begin{itemize}
\item multi-scale
sampling, and
\item robust approximation of personalized PageRanks.
\end{itemize}

}
\paragraph{Robust Approximation of PersonalizedPageRanks}

For networks with constant maximum degrees, we can simply use
  personalized PageRank
 approximation algorithms developed
  by Jeh-Widom \cite{JehW03}  or Andersen-Chung-Lang
 \cite{AndersenCL06} inside the
  multi-scale scheme to obtain an
 $\tilde{O}(n/\Delta)$ time algorithm for  {\sc SignificantPageRanks}.
However, for networks such as Web graphs and social networks that may
have nodes with large degrees, these two approaches are not
  sufficient for our needs.

We develop a new local algorithm for approximating personized PageRank
  that satisfies the desirable robustness property that our multiscale
  sample
  scheme requires.
Given $\lambda,\epsilon >0$ and a starting vertex $u$ in a network
$G=(V,E)$, our algorithm estimates each entry in the
  Pprsonalized PageRank vector defined by $u$,
  $$\textnormal{PersonalizedPageRank}(u,.)$$ to a
   $[1-\lambda,1+\lambda]$ multiplicative approximation
  around its value plus an additive error of at most $\epsilon$.
The time complexity of our algorithm is $O\left(\frac{\log^2 n
  \log(\epsilon^{-1})}{\epsilon \lambda^2}\right)$.
Our algorithm requires a careful simulation of random walks
 from the starting node $u$ to
 ensure that its complexity does not depend on the degree
 of any node.
Together with the multi-scale sampling scheme, this algorithm leads to an
 $\tilde{O}(n/\Delta)$ time algorithm for  {\sc SignificantPageRanks}.
We conclude our analysis by showing that our algorithm for solving
  {\sc SignificantPageRanks} is optimal up to a polylogarithmic
  factor.

{\blue
\begin{discussion}
While the main contribution of this paper is theoretical, that is, our
  focus is to design the first nearly optimal, local algorithm for
  PageRank approximation, we hope our algorithm or its refinements can
  be useful in practical settings for analyzing large-scale
  networks.
For example, our sublinear algorithm for {\sc SignificantPageRanks}
  could be used in Web search engines, which often need to build a
  core of  Web pages, to be later used for Web search.
It is desirable that pages in the core have high PageRank values.
These search engines usually apply crawling to discover
new significant pages and insert them to the core to
  replace existing core pages with relatively low PageRank values.
As noted already, our algorithms are local and are implementable in various
 network querying models that assume no direct global access
 to the network but allow one to generate random out-links of a given
  node as well as to uniformly at random sample nodes from the network.\
Such an implementation is desirable for
  processing large social and information networks as in the
  construction of the core pages for Web search.
We also anticipate that our algorithm for {\sc SignificantPageRanks}
and the multi-scale scheme for its matrix abstraction
will be useful for many other network analysis tasks.
\end{discussion}
}

{\blue
\paragraph{Related Work}

Our research is inspired by the body of work on
   local algorithms
  \cite{SpielmanTengLocalClustering,AndersenCL06,AndersenBCHMT08},
  sublinear-time
  algorithms \cite{RubinfeldS12}, and property testing
  \cite{Goldreich10c}
  which study algorithm design for finding relevant
  substructures or estimating various quantities of interest without
  examining the entire input. 
Particularly, we focus on identifying nodes with
  significant PageRanks and approximating personalized PageRanks
  without exploring the entire input network.
In addition, our framework is based on a combination of uniform crawling and uniform sampling of
  vertices in a graph and hence it
  can be viewed as a sublinear
  algorithm (when $\Delta = n^{\Omega(1)}$) in a rather general access
  model as discussed in \cite{RubinfeldS12}.

It is well-known that in a directed graph, high in-degree of a node does
not imply high
  PageRank for that node and vice versa.
In fact, even in real-world Web graphs, only weak
  correlations have been reported between PageRank and in-degree
  \cite{PanduranganRU06}.
One therefore needs to use methods for
  PageRank estimation that are not solely based on finding high
  in-degree nodes.}
Indeed, over the past decade, various beautiful
  methods have been developed to approximate the PageRank of all nodes.
The common thread is that they all run in time at least linear in the
  input (See \cite{Berkhin05} for a survey of results).
Perhaps the closest ones to our framework are the following
  two Monte-Carlo based approaches.
The PageRank estimation method of \cite{ALNO07}
  conducts simulation of a constant number of
  random walks from each of the nodes in the network and therefore it
  requires linear time in the size of the network.
A similar approach is analyzed in \cite{BahmaniCG10},
  where a small number of random walks
  are computed from each network node, which shows that
  a tight estimate for the PageRank of a node
  with a large enough PageRank can
  be computed from the summary statistics of these walks.
In addition, the paper shows how these estimates can be kept
  up to date, with a logarithmic factor overhead,
  on a certain type of a dynamic graph
  in which a fixed set of edges is inserted in a random order.

 {\blue  Our scheme is suitable to handle any network
  with arbitrary changes in it as well,
  including addition or removal of edges and nodes,  with the
necessary computation being performed ``on the spot'' as needed.
But in contrast to the above approaches, for $\Delta = n^{\Omega(1)}$,
our construction gives a sublinear-time algorithm for identifying all nodes whose PageRanks
are above threshold $\Delta$  and approximating
their PageRanks.}

{\blue  
We have benefited from the intuition of several previous works on
  personalized PageRank approximation.
Jeh and Widom developed a
  method based on a deterministic simulation of random walks by pushing
  out units of mass across nodes \cite{JehW03}.
Their algorithm gives an $\epsilon$-additive
 approximation with runtime cost of order of $\log n/\epsilon$
  times the maximum out-degree of a node in the network.
Andersen, Chung, and Lang \cite{AndersenCL06} provided a clever implementation of the approach of Jeh and Widom that removes the $\log n$ factor from the runtime cost, still stopping when the residual amount to push out per node is at most\footnote{Thus at termination the infinity norm of the residual vector is at most $\epsilon$, which can easily be shown to bound from above the infinity norm of the difference between the true personalized PageRank vector and the estimation computed.} $\epsilon$. We note, however, that for networks with large out-degrees, the complexity of this algorithm may not be sublinear.

Andersen {\em et al.} \cite{AndersenBCHMT08} developed a ``backwards-running''
version of the local algorithm of \cite{AndersenCL06}. Their
algorithm finds an $\epsilon$-additive approximation to the PageRank vector with
runtime proportional to $\frac{1}{\epsilon}$, times the maximum in-degree in the network, times the PageRank value. The authors show how it can be used to provide some reliable estimate to a node's PageRank: for a given $k$, with runtime proportional to $\tilde{\Theta}(k)$ times the maximum in-degree in the network (and no dependency on the PageRank value), it can bound the total contribution from the $k$ highest contributors to a given node's PageRank. However, for networks with large in-degrees, its complexity may not be sublinear even for small values of $k$. We also note that the method does not scale well for estimating the PageRank values of multiple nodes, and needs to be run separately for each target node.

}

{\blue
The problem of {\sc SignificantMatrixColumns}
  can also be viewed as a matrix sparsification or
  matrix approximation problem, where the objective is to remove all columns
  with $l_1$ norm less than $\Delta/c$ while
  keep all columns with $l_1$ norm at least $\Delta$.
To achieve time-efficiency, it is essential to allow the algorithm the
   freedom in deciding whether to keep or delete columns whose $l_1$ norm is
  in the range $[\Delta/c,\Delta]$.

While there has been a large body of work of finding a low
  complexity approximation to a matrix (such as a low-rank matrix)
  that preserves some desirable properties,
   many of the techniques developed
  are not directly applicable to our task.

First, we would like our algorithms to work even if the graph does not
have a good low rank approximation; indeed, all of our algorithms work
for any input graph. Second, our requirement to approximately preserve
$l_1$
  norm only for significant columns
  enable us to achieve $\tilde{O}(n/\Delta)$ complexity
  for any stochastic matrix, whereas all
  low-rank matrix approximations run in time
  at least linear in the number of rows and columns of the matrix in
  order explicitly reconstruct a low-rank approximation; see
  \cite{Kannan10, KannanV09} for recent surveys on low-rank approximations.

On a high level, the problem of {\sc SignificantMatrixColumns} may seem
to share some resemblance to the \textit{heavy-hitters} problem
considered in the data streaming literature \cite{CormodeM05}. In the
heavy-hitter problems, the goal is to identify all elements in a vector stream
that have value bigger than the sum of all elements. The main difficulty to
overcome is the sequential order by which items arrive and the small space
one can use to store information about them. The main technique used to
overcome these difficulties is the use of multiple hash functions which
allows for concise summary of the frequent items in the stream. However,
in {\sc SignificantMatrixColumns} we are faced with a completely
different type of constraints --- access to only a small fraction of the input
matrix (in order to achieve sublinear runtime) and having a
precision-dependent cost of matrix row-approximations.  As a result,
hashing does not seem to be a useful avenue for this goals and one needs
to develop different techniques in order to solve the problem.}

\paragraph{Organization}
In Section \ref{pagerank:sec:prelim}, we introduce some notations that
  will be used in this paper.
In Section \ref{pagerank:sec:vectorsum},
to better illustrate the multi-scale framework, we present a solution
to a somewhat simpler abstract problem that distills the computational
task we use to solve {\sc SignificantMatrixColumns}.  In particular,
we consider a blackbox model accessing an unknown vector that either
returns an exact answer or 0 otherwise.  Like the access model in {\sc
  SignificantMatrixColumns}, higher precision costs more.  In Section
\ref{pagerank:sec:multiscale}, we present our multi-scale sampling
algorithm for {\sc SignificantMatrixColumns}.  In Section
\ref{pagerank:sec:pagerank}, we address the problem of finding
significant columns in a PageRank matrix by giving a robust local
algorithm for approximating personalized PageRank vectors. The section
ends with a presentation of a tight lower bound for the cost of
solving {\sc Significant Matrix Columns} over PageRank matrices.

\section{Preliminaries}
\label{pagerank:sec:prelim}
In this section, we introduce some basic notations that we will frequently use in the paper.
{\blue
For a positive integer $n$, $[1:n]$ denotes the set of all integers $j$ such that $1 \leq j \leq n$.}
If $M\in \Reals{n\times n}$ is an $n\times n$ real matrix,
  for
 $v\in [1:n]$,
we will use $M(v,\cdot)$ and $M(\cdot,v)$ to
   denote $v^{th}$ row and the $v^{th}$ column of $M$, respectively.
We denote the sum of the column $v$ in $M$ by
$\textnormal{ColumnSum}(M,v)$.
When the context is clear we shall
  suppress $M$ in this notation and denote it by
$\textnormal{ColumnSum}(v)$.

Most graphs considered in this paper are directed.
For a given directed graph $G=(V,E)$, we usually assume $V = [1:n]$.
We use an $n \times n$ matrix $A(G)$ to denote the adjacency matrix of
$G$.
In other words, $A(i,j)=1$ if and only $(i,j)\in E$.

The PageRank vector of a graph $G$ is the (unique) stationary point of the
following equation \cite{PageBMW98,Haveliwala03}:
 $$\textnormal{PageRank}(\cdot)= \alpha \cdot \11^n + (1-\alpha)
\textnormal{PageRank}(\cdot) \cdot D^{-1}A(G),$$ where $\11^n$ is the
$n$-place row vector of all 1's, $0 < \alpha <1 $ is a teleportation
probability constant, and $D$ is a diagonal matrix with the out-degree of $v$
at entry $(v,v)$.

{\blue
Similarly, the personalized PageRank vector of $u$ in the graph $G$ is the (unique) stationary point of the
following equation \cite{Haveliwala03}:
 $$\textnormal{PersonalizedPageRank}_u(\cdot)= \alpha \cdot \11_u + (1-\alpha)
\textnormal{PersonalizedPageRank}_u(\cdot) \cdot D^{-1}A(G),$$ where $\11_u$ is the
indicator function of $u$.

Note that with the above definition of  PageRank,
the sum of the entries of the PageRank vector is normalized to $n$.
This normalization is more natural in the context of personalized PageRank than the
traditional normalization in which the sum of all PageRank entries is $1$.}

For any $x$, $\log(x)$  means $\log_2(x)$ and $\ln(x)$
  denotes the natural logarithm of $x$.

\section{Multi-Scale Approximation of Vector Sum}
\label{pagerank:sec:vectorsum} Before presenting our algorithms
for  {\sc SignificantMatrixColumns},
  we give a multi-scale algorithm for a much simpler problem that,
  we hope, captures the essence of the general algorithm.

We consider the following blackbox model for accessing an
unknown
  vector $\pp=(p_1,...,p_n)\in [0,1]^n$: We can only access the
  entries of $\pp$ by making a query of the form
 {\em vectorAccess}($i,\epsilon$).
If $p_i\geq \epsilon$, the query  {\em
vectorAccess}($i,\epsilon$)
   returns $p_i$, otherwise when $p_i<\epsilon$,
   {\em vectorAccess}($i,\epsilon$) returns 0.
Furthermore, {\em vectorAccess}($i,\epsilon$)  incurs a {\em
cost} of $1/\epsilon$. In this subsection, we consider the
following abstract problem:

\begin{quote}
{\bf {\sc VectorSum}:} {\em Given a blackbox model {\em
vectorAccess}{\rm()} for accessing
  an unknown vector $\pp = (p_1,...,p_n)\in
  [0,1]^n$, a threshold $\Delta\in [1:n]$ and a positive constant $c>1$,
  return {\textbf{PASS}} if $\sum_i p_i \geq \Delta$, return {\textbf{FAIL}} if $\sum_i p_i < \frac{\Delta}{c}$, and
  otherwise return either {\textbf{FAIL}} or {\textbf{PASS}}.
}
\end{quote}

{\blue To motivate our approach,
  before describing our multi-scale algorithm to solve this problem, let
  us first analyze the running time of a standard sampling algorithm.}
{\blue
 In such
an algorithm, one would take $h$ i.i.d. samples $s_1,
\dots, s_h$ uniformly from $[1:n]$ and query $p_{s_t}$
at some precision $\epsilon$ to obtain an estimator
$$
\frac{n}{h}\sum_{t=1}^{h} p_{s_t}\mathbf I[p_{s_t}\geq\epsilon]
$$
for the sum $\sum_ip_i$.
 The error stemming from querying at precision
$\epsilon$ would be of order $n\epsilon$, so we clearly will
have to choose $\epsilon$ of order $\Delta/n$ or smaller not to
drown our estimate in the query error, leading to a run time of
order $h n/\Delta$. The number of samples, $h$,
on the other hand, has to be large enough to guarantee
concentration, which at a minimum requires that the expectation
of the sum $\sum_{t=1}^{h} p_{s_t}\mathbf
I[p_{s_t}\geq\epsilon]$ is of order at least unity. But the
expectation of this sum is upper bounded by $(h/n)\sum
p_i$ which is of order $h \Delta/n$ in the most
interesting case where $\sum p_i$ is roughly equal to $\Delta$.
We thus need $h$ to be of order at least $ n/\Delta$,
giving a running time of order $ (n/\Delta)^2$, while we are
aiming for a sublinear running time of order $\tilde
O(n/\Delta)$. }

{\blue Our algorithm is based on a different idea by querying
$p_t$ at a different precision each time, namely, by querying
$p_{s_t}$ at precision $\epsilon_t=t/h$ in the $t^{\text{th}}$
draw, and considering the estimator
{\red
\begin{equation}
\label{Estimator}
\frac{n}h\sum_{t=1}^h \mathbf I[ p_{s_t} \geq
  \epsilon_t]
\end{equation}
for the sum $\sum_ip_i$. In expectation, this estimator
is equal to $n$ times
\begin{equation}
\label{E-Estimator}
\frac{1}h\sum_{t=1}^h \mathbf P[ p_{s_t} \geq
  \epsilon_t]
  =
  \frac{1}h\sum_{t=1}^h \mathbf P\Bigl[ p_{s_t} \geq
  \frac th\Bigr]
\end{equation}
with $s_t$ denoting an integer chosen uniformly at random from $[1:n]$.
This
is a Riemann sum approximation to the well known
expression
$$
\mathbf E[p_s]=
\int_0^1 dx \mathbf P[p_s\geq x]
$$
and differs from this integral by an error $O(\frac 1h)$.}
In the most interesting case where
$\sum_ip_i$ is of order $\Delta$, concentration again requires
$h$ to be of order at least $n/\Delta$, {\red which  also guarantees
that the error $O(1/h)$ from the Riemann sum approximation does
not dominate the expectation $\mathbf E[p_s]=\frac 1n\sum_ip_i$.}
 But now we only query
$p_s$ at the highest resolution $\epsilon_1=1/h$ once, leading
to a much faster running time.  In fact, up to log factors, the
running time will be dominated by the first few queries, giving
a running time of $\tilde O(h)=\tilde O(n/\Delta)$, as desired.
}

%
{\blue In the next section we proceed with the algorithm's
formal description and analysis.  }

\subsection{A Multi-Scale Algorithm for Approximating Vector Sum}

The following algorithm, \textit{MultiScaleVectorSum}, replaces
the standard sampling to estimate the sum $\sum_ip_i$ by a
multi-scale version which spends only a small amount of time at
the computation intensive scales requiring high precision. In
addition to the blackbox oracle $vectorAccess()$,
   this algorithm takes three other parameters:  $\Delta\in
(1,n)$ and $c>1$ as defined in {\sc VectorSum}, and a
confidence parameter $\delta \in (0,1)$:
 This algorithm uses randomization and we will show that it correctly solves {\sc VectorSum} with probability at least $1-\delta$.
{\blue Our algorithm implements the strategy discussed above except
for one modification: instead of sampling at a different
precision $\epsilon_t$ each time, we sample at each precision a constant number of times
$\tau$, where $\tau$ depends on the desired success probability,
given a total number of queries equal to $L=\tau h$, where
$h=\Theta(n/\Delta)$ with the implicit constant in the $\Theta$-symbol depending on $c$ in such a way
that it grows with $(c-1)^{-2}$ as $c\to 1$
(somewhat arbitrary, but convenient for our notation and proofs, we introduce the $c$ dependence of our constructions through
the variable
$\beta=\frac {c-1}{4c}$; in terms of this variable, we write the lower cutoff $\Delta/c$ as $\Delta(1-4\beta)$, and use
the midpoint $\Delta(1-2\beta)$ between $\Delta$ and $\Delta/c$ as the cutoff for the algorithm to decide between {\textbf{PASS}} and
{\textbf{FAIL}}).
}

\begin{algorithm}
\caption{MultiScaleVectorSum} \label{Alg:VecSum}
\begin{algorithmic}[1]
\REQUIRE {\blue $vectorAccess(\cdot,\cdot)$, threshold $\Delta
\in (1,n)$, cutoff parameter $c>1$, failure probability $\delta
\in (0,1)$.}
\STATE $\beta = \frac{c-1}{\red 4c}$;
$\tau=\lceil\log(1/\delta)\rceil$;
$h=\lceil\frac{{\red 3}n}{\Delta\beta^2}\rceil$; $L = \tau h$ \STATE
$sum = 0$. \FOR{$t=1:L$} \STATE $\epsilon_t = \frac 1h
\lceil\frac t\tau\rceil$. \STATE Let $s_t$ be an uniform random
element from $[1:n]$.
\STATE {\red $z_t = vectorAccess(s_t,\epsilon_t)$.}
\STATE $sum = sum + z_t$. \ENDFOR \IF{$sum \geq
(1-2\beta)\frac{L\Delta}{\red n}$}
   \STATE Return {\textbf{PASS}}.
\ELSE \STATE  return {\textbf{FAIL}}. \ENDIF
\end{algorithmic}
\end{algorithm}

\begin{theorem}[Multi-Scale Vector Sum]\label{theo:vecSum}
For any $\pp \in (0,1)^n$ accessible by $vectorAccess()$,
   threshold $\Delta\in (1,n)$, robust
parameter $c>1$,  and failure parameter
   $\delta \in (0,1)$,
the method \\${\rm
MultiScaleVectorSum}\left(vectorAccess(),\Delta,c,\delta\right)$
correctly solves {\sc VectorSum}  with probability at least
$(1-\delta)$ and costs
\[O\left(\frac{n}{\Delta}\left(\frac{1}{c-1}\right)^{2}\log\Bigl(\frac {n}{\Delta(c-1)}\Bigr)\log
\Bigl(\frac 2\delta\Bigr)\right).
\]

\end{theorem}
\begin{proof}
By  Steps 3-7, for any constant $c>1$, the cost of the
algorithm is
\[
\sum_{t=1}^{L}\frac 1{\epsilon_t}=\tau \sum_{i=1}^{h}\frac hi
\leq L(1+\log h)
=O\left(\frac{n}{\Delta}\left(\frac{1}{c-1}\right)^{2}\log\Bigl(\frac {n}{\Delta(c-1)}\Bigr)\log
\Bigl(\frac 2\delta\Bigr)\right).
\]
We now prove the correctness of the algorithm.

{\blue
Algorithm MultiScaleVectorSum, after the initialization Steps 1
and 2, computes the multi-scale parameters $\epsilon_t$ and
applies sampling to calculate the sum
\[
Q=\sum_{t=1}^{L} z_t=
\sum_{t=1}^L {\red \mathbf I\left[ p_{s_t}\geq \epsilon_t\right]}
\]
where $s_1,\dots,s_L$ are chosen i.i.d.~uniformly at random
from $[1:n]$. The expectation of $Q$ is easily estimated
in terms of the  bounds
\begin{align}
\mathbf E[Q]&=\frac 1n\sum_{k=1}^n \sum_{t=1}^{L} {\red\mathbf I \left[\epsilon_t \leq p_k\right]}
=
\frac 1n\sum_{k=1}^n \sum_{t=1}^{ L}{\red \mathbf I\left[\left\lceil t/\tau\right\rceil\leq hp_k\right]}
\notag
\\&=
\frac\tau n \sum_{k=1}^n \sum_{i=1}^{h}
{\red \mathbf I\left[ i\leq {hp_k}\right]
=\frac\tau n \sum_{k=1}^n\lfloor hp_k \rfloor}
\leq
\frac \tau n\sum_{k=1}^n  hp_k
=
\frac {L}{n}\sum_{k=1}^n  p_k
\label{eqn:Qub}
\end{align}
and
{\red
\begin{equation}
\label{eqn:Qlb}
\mathbf E[Q]\geq
\frac \tau n\sum_{k=1}^n \Bigl( hp_k -1\Bigr)=
\frac {L}{n}\sum_{k=1}^n  p_k -\tau.
\end{equation}
We thus use $\frac{ n}{L} Q$}
as an estimate
of $\sum_{k=1} ^np_k$ when we decide on whether to output {\textbf{PASS}}
  in Step 9.}

Assume first that $\sum p_k\geq \Delta$.  Since
$\tau\leq{\red\beta^2\frac {L\Delta}{3n}}\leq \beta\frac
{L\Delta}{n}$, we then have
\[
\mathbf E[Q]\geq \frac {L\Delta}{n} -\tau
\geq
(1-\beta){\red \frac{L\Delta}{n}},
\]
implying that
\[
(1-\beta)\mathbf E[Q]\geq (1-2\beta){\red \frac{ L\Delta}{n}.}
\]
This allows us to use the
multiplicative Chernoff bound in the form of Lemma
\ref{lemma.multchernoff} to conclude that
\[
{\mathbf P\mathbf r}\left[Q\leq (1-2\beta){\red \frac{ L\Delta}{n}}\right]
\leq
{\mathbf P\mathbf r}\left[
Q\leq (1-\beta)\mathbf E[Q]
\right] \leq
\exp\left(-\frac{\beta^2}2 \mathbf E[Q]\right)
\leq \exp\left(-{\red\frac 38}\frac{\beta^2L\Delta}{ n}\right)\leq \delta,
\]
where we used $\beta\leq {\red 1/4}$ in the last step.

On the other hand, if
 $\sum p_k\leq \Delta/c=(1-{\red 4}\beta)\Delta$, we  bound
\[
\mathbf E[Q]\leq {\red \frac { L\Delta}{n}}(1-{\red 4}\beta)
\]
which in turn implies that
\[
(1+{\red 2}\beta) \mathbf E(Q)\leq (1-2\beta){\red \frac{ L\Delta}{n}}.
\]
Using the multiplicative Chernoff bound in the form
\ref{lemma.multchernoff} (part 3), this gives
\[
\begin{aligned}
Pr\left[Q\geq (1-2\beta){\red \frac{ L\Delta}{n}}\right]
&\leq
 \exp\left(-\beta^2{\red \frac{ L\Delta}n}\frac{1-2\beta}{1+{\red 2}\beta} \right)
 \leq
  \exp\left(-\frac{\beta^2  L\Delta}{{\red 3}n} \right)
 \leq \delta.
\end{aligned}
\]
where we again used $\beta\leq {\red 1/4}$.

Thus, ${\rm
  MultiScaleVectorSum}\left(vectorAccess(),\Delta,c,\delta\right)$
correctly solves {\sc VectorSum} with probability at least
$1-\delta$.

\end{proof}
\vspace{0.2in}

\section{Multi-Scale Matrix Sampling}
\label{pagerank:sec:multiscale}

In this section, we consider {\sc SignificantMatrixColumns} in
  a slightly more
  general matrix access model than what we defined in Sections 1 and
  3.
The extension of the model is also needed in our PageRank
approximation algorithm, which we will present in the next section.

\subsection{Notation: Sparse Vectors}
To better specify this model and the subsequent algorithms, we first
  introduce  the notation of  {\em sparse vector}
  introduced by Gilbert, Moler, and
  Schreiber \cite{sparseMatlab} for Matlab.
Suppose $\aa = (a_1,...,a_n) \in \Reals{n}$
  is a vector.
Let $\nnz(\aa)$ denotes the number of nonzero elements in $\aa$.
Let $\sparse(\aa)$ denote the {\em sparse form} of vector $\aa$
   by ``squeezing out'' any zero elements in $\aa$.
Conceptually, one can view $\sparse(\aa)$ as a list of $\nnz(\aa)$
index-entry pairs, one for each nonzero element and its index in
$\aa$.
For example, we can view $\sparse([0,0.3,0.5,0,0.2])$ as
  $\left((2,0.3),(3,0.5), (5,0.2)\right).$

A sparse vector can be easily implemented using a binary search tree \footnote{For average case rather than worst-case guarantees, a hash table is a typical implementation choice.}. Throughout the paper we shall make use of the following simple proposition:

\begin{proposition} \label{prop:sumofsparsevec}
For $\aa,\bb\in \Reals{n}$, $\aa + \bb$ can be
implemented in time $O(\nnz(\bb) \cdot \log{n})$ saving the result in the data structure of $\aa$.
\end{proposition}
\begin{proof}
{\blue
Each sparse vector can be implemented as a balanced binary search tree,
  where the index of an entry serves as the entry's key.
When performing the addition, we
  update the binary search tree of $\aa$ by inserting one by one the
  elements of $\bb$ into it (and updating existing entries whenever
  needed).
By the standard theory of binary search trees, each such insertion
  operation takes $O(\log{n})$ time.  }
\end{proof}

In the rest of the paper, without further elaboration, we
  assume all vectors are expressed in this sparse form.
We also adpot the following notations: let $\sparse([\ ])$ denote the
all zero's vector in the sparse form, and for any $i \in [1:n]$ and $b\in
\Reals{}-\{0\}$, let $\sparse(i,b)$ denote the sparse vector with only
one nonzero element $b$ located in the $i^{th}$ place in the vector.
In addition, we will use the following notation:
For two vectors $n$-place vectors
  $\aa = (a_1,...,a_n)$ and $\bb=(b_1,...,b_n)$,
  and parameters $\epsilon \in \Reals{}$ and
$C > 0$, we use $\aa \leq C\cdot \bb +\epsilon$ to denote
 $a_i \leq C\dot b_i + \epsilon$,   $\forall i\in [1:n]$.

\subsection{The Matrix Access Model}
In the model  that we will consider in the rest of this section,
  we can access an unknown
  $n\times n$ {\em right-stochastic matrix}
  $M =\left(m_{i,j}\right)$ using queries of the form
    {\em matrixAccess}($i,\epsilon, \lambda, p$),
    where $i \in [1:n]$ specifies a row, $\epsilon \in (0,1]$
      specifies a required additive precision, $\lambda \in (0,1]$
        specifies a multiplicative precision, and $p\in (0,1]$
          specifies the probability requirement.
{\blue This query will return a sparse vector $\approxmm_i =
\sparse([\approxm_{i,1},...,\approxm_{i,n}])$ such that
}
\begin{itemize}
\item with probability at least $1-p$,
\begin{equation}
\label{equ:approx-bd}
(1-\lambda)\cdot \mm_i -\epsilon\leq
  \approxmm_i \leq (1+\lambda)\cdot \mm_i + \epsilon,
  \end{equation}
where $\mm_i = M(i,\cdot)$ denotes the $i^{th}$ row of matrix $M$, and
\item {\blue with probability at most $p$ (the query may fail),
   $\approxmm_i$ could be an arbitrary sparse vector}.
\end{itemize}
We refer to this blackbox model as the {\em probabilistic
  sparse-and-approximate
  row-access model with additive/multiplicative errors}.
For constant integers $c_1, c_2,c_3, c_4 > 0$, we say that $matrixAccess$ is
 an $(c_1,c_2,c_3,c_4)$-SARA model if
 for all $i \in [1:n]$, $\epsilon \in (0,1)$, $\lambda \in (0,1)$,
  and $p\in (0,1)$, both the cost of calling
$\approxmm_i = ${\em matrixAccess}($i,\epsilon, \lambda, p$) and
 $\nnz(\approxmm_i)$ are bounded from above by
$$
 c_1 \left(\frac{1}{\lambda}\right)^{c_2}\left(\frac{\log^{c_3}(1/\epsilon)}{\epsilon}\right) \left(\log^{c_4}{n}\right) \log (1/p).
$$

\subsection{The Matrix Problem}

In this section, we give a solution
to the following abstract problem.
\begin{quote}
{\bf {\sc SignificantMatrixColumns}:}
{\em Given an $n\times n$ {\em right-stochastic matrix} $M$ in the
   $(c_1,c_2,c_3,c_4)$-SARA model, a threshold $\Delta$ and a positive constant $c>1$,
  return a sparse vector $cSum$ with the
  property that for all $j\in [1:n]$, if
  $\textnormal{ColumnSum}(M,i)\geq \Delta$, then $cSum(j)\neq 0$ and if
    $\textnormal{ColumnSum}(M,i)< \Delta/c$, then $cSum(j) = 0$.}
\end{quote}

\subsection{Understanding the Impact of Additive/Multiplicative Errors}\label{sec:errors}

Our algorithm for {\sc SignificantMatrixColumns} is straightforward.
At a high level, it simultaneously applies Algorithm~\ref{Alg:VecSum} to all columns of the unknown matrix.
It uses a sparse-vector representation for efficient bookkeeping of the
  columns with large sum according to the sampled data.
Our analysis of this algorithm is similar to the one presented that in Theorem
\ref{theo:vecSum} for {\sc VectorSum} as we can
  use the union bound over the columns to
  reduce the analysis to a single column.
The only technical difference is the handling of the
  additive/multiplicate errors.

To understand the impact of these errors, we consider a vector
$\pp = (p_1,...,p_n)\in {\red [0,1]^n}$ and chose $\epsilon_t$, $t=1,\dots,L$ as in Algorithm~\ref{Alg:VecSum}.
Fix $\phi,\lambda\in (0,1/2)$, and suppose
that we access $p_i$ with multiplicative error $\lambda$ and additive error
$\phi\cdot \epsilon_t$.
We will show that if this returns a
number
{\red $\approxp_i\geq \epsilon_t$, the actual value of $p_i$ is at least
$\rho\epsilon_t$,} where $\rho=1-\lambda-\phi$.
To see this, we bound
\[
p_i\geq
(1+\lambda)^{-1}(\approxp_i-\phi\cdot\epsilon_t)
\geq
(1+\lambda)^{-1}(1-\phi)\epsilon_t
.\]
Since $(1-\phi)/(1+\lambda)\geq (1-\lambda-\phi)$, this implies
{\red $p_i\geq \rho\epsilon_t$}, as desired.

In a similar way, it is easy to see that
{\red $p_i\geq\rho^{-1}\epsilon_t$ implies that $\approxp_i\geq\epsilon_t,$.  Indeed,
if $p_i\geq\rho^{-1}\epsilon_t$ then}
\[
 \approxp_i
 \geq
(1-\lambda) p_i-\phi\cdot\epsilon_t
 \geq
\left(\frac{1-\lambda}{1-\lambda-\phi}-\phi\right)\epsilon_t.
\]
The lower bound is clearly larger than $\epsilon_t$, , showing that
{\red $\approxp_i\geq\epsilon_t$.}

For $s_1\dots,s_{L}\in [1:n]$, the sum
\begin{equation}
\label{eqn:tildeQ}
\tilde Q=
\sum_{t=1}^{ L}{\red  \mathbf I[ \approxp_{s_t}\geq \epsilon_t]}
\end{equation}
can therefore be bounded from below and above by
\begin{equation}
\label{eqn:Q-+}
Q_-=
\sum_{t=1}^L {\red \mathbf I[ p_{s_t} \geq\rho^{-1}\epsilon_t ]}
\quad\text{and}\quad
Q_+=
\sum_{t=1}^L {\red \mathbf I[  p_{s_t} \geq\rho\epsilon_t ]},
\end{equation}
respectively:
\begin{equation}
Q_-\leq \tilde Q\leq Q_+.
\label{eqn:tildeQ-bd}
\end{equation}
{\red Finally, we also note that if we access
$p_i$ with multiplicative error $\lambda$ and additive error
$\phi\cdot \epsilon_t$, then this a returns a number
which is never larger than $\rho^{-1}$.  Indeed, this follows by bounding
$\tilde p_i$ by $1+\lambda +\phi\cdot \epsilon_t\leq 1+\lambda +\phi\leq \rho^{-1}$.
}

\subsection{A Multi-Scale Algorithm}
{\blue
In this section we present the multi-scale algorithm in full details and proceed with an analysis of its runtime and correctness.
The algorithm is essentially an extension of
Algorithm~\ref{Alg:VecSum}, applying the VectorSum algorithm to all
columns in parallel. As now the call to vectorAcesss has been replaced
by a combined additive-multiplicative method, the constant $\beta$ is
set to a slightly smaller value than in Algorithm~\ref{Alg:VecSum}. In
addition to the constants $\beta,\tau,h, L$ that are used in
Algorithm~\ref{Alg:VecSum}, we also have the constant $\lambda$ for
the value of multiplicative approximation needed and $\phi$ for the
additive-approximation needed. Lastly, $p$ is the wanted success
probability of the row approximation procedure (matrixAccess) invoked
throughout the algorithm.
{\blue We note that these constants are defined to allow complete and rigorous analysis of our algorithm and its correctness.}
As the multi-scale algorithm will essentially be implementing Algorithm~\ref{Alg:VecSum} over all columns, we will need a method that can return all elements in a row that fall within a certain bin; we call it the \textit{rangeIndicator} method.}

$\rangeIndicator()$: for a sparse vector $\aa$, and $l,u\in\Reals{}$ such
that $l<u$,
$\bb = \rangeIndicator(\aa, l,u)$
returns a
sparse vector $\bb$ such that for all $i\in [1:n]$,
$$\bb(i) = \left\{\begin{array}{ll} {1}\ \ \ \  & \mbox{if } l\leq \aa(i)\leq u\\
                                     0 & \mbox{otherwise}
                 \end{array}
           \right.
$$
For example, $\rangeIndicator(\sparse([0,0.3,0.5,0,0.2]), 0.1,0.3)$ returns
  the sparse form of $[0,{1},0,0,{1}]$.
We shall use the following simple proposition:
\begin{proposition}
\label{prop:indicator}
   {$\rangeIndicator(\aa, l,u)$} takes $O(\nnz(\aa){\log n})$ time.
\end{proposition}
\begin{proof}
The sparse vector $\nnz(\aa)$ is implemented using a binary search tree; one can therefore scan its content using, say, an inorder scan and insert each element in the range $[l,u]$ to a sparse vector $\bb$, initially empty.
The inorder scan costs $O(\nnz(\aa))$ time and each insertion into $b$ costs $O({\log n})$ time, giving the desired result.
\end{proof}

\begin{algorithm}
\caption{MultiScaleColumnSum} \label{multiScaleColumnSum}
\begin{algorithmic}[1]
\REQUIRE {\blue $matrixAccess(\cdot,\cdot,\cdot,\cdot)$,\ \  threshold $\Delta\in (1,n)$,\ \  cutoff $c>1$,\ \  failure probability $\delta \in (0,1)$.}
\STATE $\beta = \frac{c-1}{\red 5c}$;
\ \  $\tau=\lceil\log(2n/\delta)\rceil$;
\ \ $h=\lceil\frac{{\red 3}n}{\Delta\beta^2}\rceil$;
\ \ $L = \tau h$;
\ \ $p = \delta/(2L)$;
\ \ $\lambda = \beta/{\red 2}$; \ \  $\phi=\beta/{\red 2}$;
\ \ $\rho=1-\lambda-\phi$.
\STATE  $cSum = \sparse([\ ])$.
\FOR{$t=1:L$}
\STATE $\epsilon_t = \frac 1h\lceil  \frac t\tau \rceil$.
    \STATE Let $s_t$ be an uniform random element from $[1:n]$;
    \ \ $q_t=matrixAccess(s_t,\phi\cdot\epsilon_t,\lambda,p)$.
    \STATE   $\zz_t =
    \rangeIndicator\left(\qq_t,
  \epsilon_t,{\red\rho^{-1}}\right)$.
\STATE $cSum = cSum + \zz_t$.
\ENDFOR
\STATE $cSum = \rangeIndicator\left(cSum, (1-2\beta){\red \frac{L\Delta}{n}},L\right)$;
\STATE Return $cSum$.
\end{algorithmic}
\end{algorithm}

We are now ready to state our main theorem.
\begin{theorem}[Multi-Scale Column Sum]\label{theo:columnSum}
For any {\em right-stochastic matrix} $M$ accessible by $matrixAccess$,
  threshold  $\Delta\in (1,n)$, robust parameter {$c>1$,}
  and failure parameter   $\delta \in (0,1)$,
  with probability at least $(1-\delta)$,
 $$cSum = {\rm
   MultiScaleVectorSum}\left(vectorAccess(),\Delta,c,\delta\right).$$
correctly solves {\sc SignificantMatrixColumns}.

Furthermore, if $matrixAccess$ is
 an $(c_1,c_2,c_3,c_4)$-SARA model, then the cost \\ of
${\rm
   MultiScaleVectorSum}\left(vectorAccess(),\Delta,c,\delta\right)$ is
\[
O\Biggl(c_1 \left(\frac n\Delta\right)\left(\frac{1}{c-1}\right)^{c_2+3}\log^{c_3+2}\Bigl(\frac 1{c-1}\Bigr)
\log^{c_3+c_4+3}{n}
\log^2\Bigl(\frac 2\delta\Bigr)\Biggr).
\]

\end{theorem}

\begin{proof}
The cost of the algorithm is dominated by the sparse matrix operations in line 5-7, plus the cost of the last
operation in line 9.  Using  our access model together with Propositions~\ref{prop:sumofsparsevec} and \ref{prop:indicator},
the cost of the steps in line 5-7 at time $t$ are of order
\[
O\Biggl(c_1\left(\frac{1}{\beta}\right)^{c_2} \frac{\log^{c_3}(\frac 1{\beta\epsilon_t})}{\beta\epsilon_t}
 \,\log^{c_4+1}{n}\, \log (2L/\delta)\Biggr)
\leq
\]
\[
O\Biggl(c_1\frac h{\lceil t/\tau\rceil}\log^{c_3}h\,\left(\frac{1}{\beta}\right)^{c_2+1}\log^{c_4+1}{n}
\log (\frac {2n}{\Delta\beta\delta})\Biggr).
\]
Note that this includes the extra factor of $\log n$ from Proposition~\ref{prop:sumofsparsevec}, a factor which
 is absent in the sparseness of $\qq_t$ and $\zz_t$.

Summing over $t$ gives a running time of order
\[
\begin{aligned}
O\Biggl(c_1 L&\log^{c_3+1}h\left(\frac{1}{\beta}\right)^{c_2+1}
\log^{c_4+1}{n}
\log (\frac {2n}{\Delta\beta\delta})\Biggr)
\\
&=O\Biggl(c_1 \left(\frac n\Delta\right)\log^{c_3+1}\Bigl(\frac n{\Delta\beta}\Bigr)
\left(\frac{1}{\beta}\right)^{c_2+3}
\log^{c_4+1}{n}
\log \Bigl(\frac {2n}{\Delta\beta\delta}\Bigr)\log\Bigl(\frac 2\delta\Bigr)\Biggr)
\\
&=
O\Biggl(c_1 \left(\frac n\Delta\right)\left(\frac{1}{\beta}\right)^{c_2+3}\log^{c_3+2}\Bigl(\frac 1{\beta}\Bigr)
\log^{c_3+c_4+3}{n}
\log^2\Bigl(\frac 2\delta\Bigr)\Biggr).
\end{aligned}
\]

To estimate the cost of the last step of the algorithm, we bound the
sparseness of $cSum$ at the completion of the FOR loop by
$\nnz(cSum)\leq \sum_t\nnz(\zz_t)$ and then apply
Proposition~\ref{prop:indicator} once more, giving a cost which of
the same order as the total cost of the algorithm accrued up to this step.

To prove the correctness of the algorithm, we first
note that with probability at least $(1-p)^L\geq 1-pL$, each of the $L$ calls of
$matrixAccess$ in line 5 will
return a sparse vector obeying the bound
\eqref{equ:approx-bd}.  Next, we apply the union bound to reduce the focus of the analysis to a
  single column:
\begin{eqnarray*}
& \prob{}{\mbox{{\sc MultiScaleColumnSum} is unsuccessful}} \ \
  \leq \quad\quad\quad \quad\quad\quad\\
& \mbox{\quad }\quad \quad \quad \quad
\sum_{i=1}^n
 \prob{}{\mbox{{\sc MultiScaleColumnSum} is unsuccessful on column
     $i$}}.
\end{eqnarray*}
When considering column $i$,
  we now let $\pp = (p_1,...,p_n)^T= M(\cdot,i)$, the $i^{th}$ column
  of $M$.
In other words, $p_j = m_{i,j}$ for all $j \in [1:n]$.
Note that the $i^{th}$ entry of $cSum$ after step 8 is of the form \eqref{eqn:tildeQ}.
Taking into account the bound \eqref{eqn:tildeQ-bd}, our proof will be very similar to that of Theorem~\ref{theo:vecSum}.

We first consider the case
that $\sum_i p_i\geq \Delta$ in which case we bound
{\red
\[
\begin{aligned}
\mathbf E[Q_-]&\geq
\frac{L\rho}{n}
\sum_kp_k -\tau
\geq \frac{\Delta L}{n}\left[\rho -\frac{\beta^2}3\right]
=\frac{\Delta L}{n}\left[1-\beta-\frac{\beta^2}3\right].
\end{aligned}
\]}
Multiplying both sides by $(1-\beta)$,
we obtain
\[
(1-\beta)\mathbf E[Q_-]\geq
{\red \frac{\Delta L}{n}}\left(1-2\beta\right).
\]
Combined with the bound \eqref{eqn:tildeQ-bd} and the multiplicative Chernoff bound (Lemma A.1), this shows that conditioned on
$matrixAccess$
returning a sparse vector obeying the bound
\eqref{equ:approx-bd} in each instance in line 5, we get
\[
\begin{aligned}
\mathbf P\mathbf r\left\{cSum(i)\leq (1-2\beta){\red \frac{\Delta L}{n}}\right\}
\leq \exp\left(-\frac{\beta^2}2 \mathbf E[Q_-]\right)
\leq \exp\left(-{\red\frac 38}\frac{\beta^2L\Delta}{ n}\right)\leq \frac\delta{2 n} .
\end{aligned}
\]

In a similar way, if $\sum_kp_k\leq \Delta/c=\Delta(1-{\red 5}\beta)$, we bound
\[
\mathbf E[Q_+]\leq {\red \frac{\Delta L}{n\rho}(1-5\beta)
= {\red \frac{\Delta L}{n}}\frac{1-5\beta}{1-\beta},}
\]
implying that
\[
(1+{\red 2}\beta)\mathbf E[Q_+]\leq {\red \frac{\Delta L}{n}}(1-2\beta)
\]
and hence
\[
\begin{aligned}
\mathbf P\mathbf r\left\{cSum(i)\geq (1-2\beta){\red \frac{\Delta L}{n}}\right\}
\leq \exp\left(-{\red \beta^2 \frac{\Delta L}{n}\frac{1-2\beta}{1+2\beta}}\right)
\leq \exp\left(-\frac{\Delta L\beta^2}{{\red 3}n}\right)\leq \frac\delta{2 n},
\end{aligned}
\]
again conditioned on $matrixAccess$
returning a sparse vector obeying the bound
\eqref{equ:approx-bd} in each instance in line {\red 5}.

Thus the total failure probability is at most $Lp + n\frac{\delta}{2n}=\delta$, as desired.

\end{proof}

\section{Identifying Nodes with Significant PageRank}
\label{pagerank:sec:pagerank}

{\blue
\subsection{Robust Approximation of Personalized PageRanks}
}

{\blue We now present our main subroutine for {\sc
    SignificantPageRanks} which, we recall, is {\red addressing the following problem:}
Given a directed graph $G =(V,E)$,
a threshold value $1\leq \Delta \leq |V|$
and a positive constant  $c>1$,
compute a subset $S\subseteq V$ with the
property that $S$ contains all vertices of
PageRank at
least $\Delta$ and
no vertex with PageRank less than $\Delta/c$.}

{\blue
Let $PPR$ denote the personalized PageRank Matrix of $G$
  defined in the
Introduction, where {\red we} recall that $PPR(i,j)$
  is equal to the personalized PageRank contribution of node $i$
  to node $j$ in $G$.
Under this notation, the {\sc SignificantPageRanks}
  can be viewed as a {\sc SignificantMatrixColumns} problem,
  if we can develop an efficient procedure for accessing the rows of $PPR$.
This procedure, which we refer to as
\textit{PPRmatrixAccess()},
 takes a row number $i$, an additive precision parameter
 $\epsilon$, a multiplicative precision parameter $\lambda$ and
 success probability $p$, and returns a sparse vector $\approxmm_i =
 \sparse([\approxm_{i,1},...,\approxm_{i,n}])$ such
 that
\begin{itemize}
\item   with probability at least $1-p$,
$$(1-\lambda)\cdot \mm_i -\epsilon\leq
  \approxmm_i \leq (1+\lambda)\cdot \mm_i + \epsilon,$$
where $\mm_i = PPR(i,\cdot)$, and
\item with probability at most $p$,  $\approxmm_i$ can be any sparse vector.
\end{itemize}
}

{\blue
Our algorith for \textit{PPRmatrixAccess()}
  use the following key observation that
  connects personalized PageRank with the hitting probability
  of a Markov model.
\begin{observation}
$PPR(v,j)$ is equal to the success probability
  that a random walk starting at $v$ and
  independently terminating at each time step with
  probability $\alpha$, hits $j$ just before termination.
\end{observation}
\begin{proof} 
Let $1_{v}$ be the indicator vector of $v$.
Solving the
  system given by
 $$\textnormal{PersonalizedPageRank}(v,\cdot)= \alpha  \textbf{1}_{v}
  + (1-\alpha)  \textnormal{PersonalizedPageRank}(v,\cdot)  D^{-1}A,$$
one obtains
$$\textnormal{PersonalizedPageRank}(v,\cdot)
= \alpha \textbf{1}_{v} (I-(1-\alpha)D^{-1}A)^{-1}= \alpha  \textbf{1}_{v} \sum_{i=0}^{\infty}{ ((1-\alpha)D^{-1}A)^{i} } .$$
The observation then follows directly from the last equation.
\end{proof}
}
{\blue
Our algorithm for \textit{PPRmatrixAccess} given below
  conduct{\red s} a careful simulation of such restarting random walks.
As such it only needs an oracle access to a random out-link of a given node.
}

\begin{algorithm}
\caption{PPRmatrixAccess} \label{approxrow}
\begin{algorithmic}[1]
\REQUIRE  {\blue node $v$, additive approximation $\epsilon$, multiplicative approximation $\lambda$.}

\STATE $cSum = \sparse([\ ])$.
\STATE Set $length =  \lceil \log_{\frac{1}{(1 - \alpha)}}(\frac{4}{\epsilon})  \rceil $.
\STATE Set $r = \lceil \frac{1}{\epsilon \lambda^2}  \cdot 4 \ln(n/p) \rceil$.
\FOR{ $r$ rounds}
      \STATE Run one realization of a restarting random walk from $v$. Artificially stop the walk after $length$ steps if it has not terminated already.
      \IF{the walk visited a node $j$ just before making a termination step}
           \STATE $cSum = cSum + Sparse(j,1/r)$ ~~~~//namely, add $1/r$ to $j$'s value.
      \ENDIF
\STATE Return $cSum$.
\ENDFOR
\end{algorithmic}
\end{algorithm}

\begin{theorem}
For any node $v$, values $0 < \epsilon <1$, $0 < \lambda <1$, $0 <\alpha < 1$, and success probability $0 < p <1$, $\textnormal{PPRmatrixAccess}(v,\epsilon,\lambda,p)$ is a $\large( 10 \max \{\log^{-1}(\frac{1}{1-\alpha}) ,1 \},2,1,2 \large)$-SARA model. In particular, its runtime is upper bounded by
$$O \left ( \frac{\ln^2(n) \ln(1/p) \log(\epsilon^{-1})}{\epsilon \lambda^2} \right ) .$$
\end{theorem}

\begin{proof}
We start by analyzing the runtime guarantee.  The algorithm performs
$\lceil \frac{1}{\epsilon \lambda^2} \cdot 4 \ln(n/p) \rceil $ rounds
where at each round it simulates a random walk with termination
probability of $\alpha$ for at most $length$ steps. Each step is
simulated by taking uniform sample ('termination' step) with
probability $\alpha$ and by choosing a random out-link with
probability $1-\alpha$.  The update of $cSum$ in line $7$ takes at
most $\log{n}$ (see proposition~\ref{prop:sumofsparsevec}).
Thus the total number of queries used is
\[
\left\lceil \frac{4\ln(n/p)}{\epsilon \lambda^2} \right\rceil  \cdot \left \lceil \log_{\frac{1}{(1 - \alpha)}}\left(\frac{4}{\epsilon}\right) \right\rceil  \log(n) \le
 \left\lceil \frac{4\ln(n/p)}{\epsilon \lambda^2}\right \rceil   \cdot \left \lceil \frac{\log(\frac{4}{\epsilon})}{\log(\frac{1} {1-\alpha})} \right\rceil \log(n) \le \]
 \[
 (8+2) \max \left \{ \log^{-1}(\frac{1}{1-\alpha}), 1 \right \} \frac{\ln^2(n) \log(1/p) \log(\epsilon^{-1})}{\epsilon \lambda^2}.
\]

We now prove the guarantees on
  the returned vector $cSum$ (line $9$ in the algorithm).
Given a node $j$, denote by $p_k(v,j)$ the contribution to $j$ from restarting walks originating at $v$ that are of length at most $k$, namely,
\[ p_k(v,j) = \alpha \textbf{1}_{v} \sum_{i=0}^{k}{ (1-\alpha)D^{-1}A)^{i}} .\]

We ask how much is contributed to $j$'s entry from restarting walks of length bigger or equal to $k$. The contribution is at most $(1-\alpha)^{k}$ since the walk needs to survive at least $k$ consecutive steps.
Taking $(1-\alpha)^{k} \le \frac{\epsilon}{4}$ will guarantee that at most $\frac{\epsilon}{4}$ is lost by only considering walks of length smaller than $k$, namely:
\[ PPR(v,j) -\frac{\epsilon}{4} \le  p_k(v,j) \le PPR(v,j) .\]
For this to hold it suffices to take $k = \lceil \log_{\frac{1}{(1-\alpha)}}{(\frac{4}{\epsilon})} \rceil$,
{\red the value the parameter  $length$ is set to in step 2.}

Next, the algorithm computes an estimate of $p_k(v,j)$ by realizing walks of length at most $k$. This is the value
of $cSum$ at index $j$ returned by the algorithm. Denote this by $\hat{p}_k(v,j)$.
The algorithm computes such an estimation (in line 7) by taking the average number of hits over $r$ trials (adding $1/r$ per hit).

Now, if $ PPR(v,j) \ge \frac{\epsilon}{2}$ then $p_k(v,j) \ge \frac{\epsilon}{4}$ and by the multiplicative Chernoff bound (Lemma \ref{lemma.multchernoff}),
$$
Pr \left(\hat{p}_k(v,j) > (1+\lambda) p_k(v,j) \right) \le \exp(- \ln(n/p) )$$
and
$$Pr \left(\hat{p}_k(v,j) < (1-\lambda) p_k(v,j) \right) \le \exp(- \ln(n/p) ) .
$$

By the union bound we can conclude that with probability $1-{\frac{2p}{n}}$,
$$(1-\lambda)(PPR(v,j) - \frac{\epsilon}{4}) \le \hat{p}_k(v,j) \le (1+\lambda)PPR(v,j) .$$

Similarly, if $PPR(v,j) < \frac{\epsilon}{2}$ then  $p_k(v,j) < \frac{\epsilon}{2}$ and by the multiplicative Chernoff bound (Lemma \ref{lemma.multchernoff}, part 3), $$Pr \large \left(\hat{p}_k(v,j) > (1+\lambda)\frac{\epsilon}{2} \right) \le \exp(-\ln(n/p) ) = p/n .$$

As $\lambda < 1$ we therefore have $ 0 \leq \hat{p}_k(v,j) \leq \epsilon$ with probability at least $1-p/n$.
And as $PPR(v,j) < \frac{\epsilon}{2}$ we clearly have, with probability $1-p/n$,
$$(1-\lambda) PPR(v,j) - \epsilon  \leq  \hat{p}_k(v,j) \leq  (1+\lambda)PPR(v,j) + \epsilon ,$$ as needed.

By the union bound, the complete claim holds with probability at least $1-p$.
\end{proof}

\subsection{A Tight Lower Bound for Solving the {\sc SignificantPageRanks} Problem}
In this subsection, we present a corresponding lower bound for
identifying all nodes with significant PageRank values.  Our lower
bound holds under the stringent model where one can access any node of
interest in the graph in one unit of cost and that the PageRank of the
node accessed is given  for free.  We call such a model the
\textit{strong query model}.
{\blue We first give a lower bound to illurstrate the challenge for
  identifying nodes with significant PageRanks, even in graphs where
  there is only one significant node.
}
{\blue
We then show that for any integral threshold $\Delta$ and
  precision $c$ there are instances where the output size of {\sc
    SignificantPageRanks} is $\Omega({n}/{\Delta})$.
Clearly, this also
  serves as a lower bound for the runtime of any algorithm that solves
  the {\sc SignificantPageRanks} problem, regardless of the
  computational model used to compute the required output.
We note that the
  runtime of our algorithmic solution to {\sc SignificantPageRanks} is
  at most only a small polylogarithmic factor away from this bound.}

For clarity of exposition we present our lower bounds for
$\alpha = 0.5$. Similar lower bounds hold for any fixed $0 < \alpha <1$.

\begin{theorem}[Hardness for Identifying One Significant Node]\label{lowerboundpagerank}
Let $\alpha = 0.5$. For $n$ large enough, any algorithm making less than $\frac{n}{6\Delta}$ queries in the strong query model on graphs on $n$ nodes
and threshold $\Delta \leq \frac{n}{9}$, would fail with probability at least $1/e$ to find a node with PageRank at least $\Delta$, on at least one graph on $n$ nodes.
\end{theorem}
\begin{proof}
{\blue
The proof will apply Yao's Minimax
  Principle for analyzing randomized algorithms~\cite{Yao77},
  which uses the average-case complexity of the deterministic
  algorithms to derive a lower bound on the randomized algorithms for
  solving a problem.

Given positive integers $n$ and $\Delta \leq \frac{n}{9}$, we
construct a family {\red $\mathcal F$} of undirected graphs on $n$ nodes by taking a cycle
subgraph on $n-d-1$ nodes and an isolated star subgraph on the {\red remaining}
$d+1$ nodes, where we set $d = 3 \Delta -1$. To complete the
construction we take a random labeling of the nodes.  See Figure
\ref{fig:pageranklower} for an illustration.

Let $A$ be a deterministic algorithm for the problem.
  We shall analyze the behavior of $A$ on
 {\red a uniformly random graph from $\mathcal F$.}

First, by solving the PageRank equation system it is easy to check
that each node on the cycle subgraph has PageRank value of $1$, the
hub of the subgraph has PageRank $\frac{d}{3}+\frac{2}{3}$, and a leaf
of the star subgraph has PageRank $\frac{2}{3}+\frac{1}{3d}$.
The only node with PageRank at least $\Delta$ is the hub of the star subgraph.

Let $T$ be the number of queries the algorithm make.
The probability that none of the nodes of the star subgraph
are found after $T$ queries by $A$ is at least
\[
(1-\frac{d+1}{n})^T \geq \exp(-2T\frac{d+1}{n}) \ge \exp(-1) ,\]
for
$T \leq \frac{n}{6\Delta} = \frac{n}{2(d+1)}$. Here we used the fact that $1-x \geq \exp(-2x)$, for $ 0 \leq x \leq 1/3$.

We define the cost of the algorithm as $0$ if it has found a node of
Pagerank at least $\Delta$ and $1$ otherwise.
Note that the cost of an algorithm equals its probability of failure.
Then by Yao's Minimax Principle, any randomized algorithm that makes
  at most $\frac{n}{6\Delta}$ queries will have an
expected cost of at least $1/e$, i.e., a failure probability of at
least $1/e$ on at least one of the inputs.  }
\begin{figure}
\begin{center}
\epsfig{file=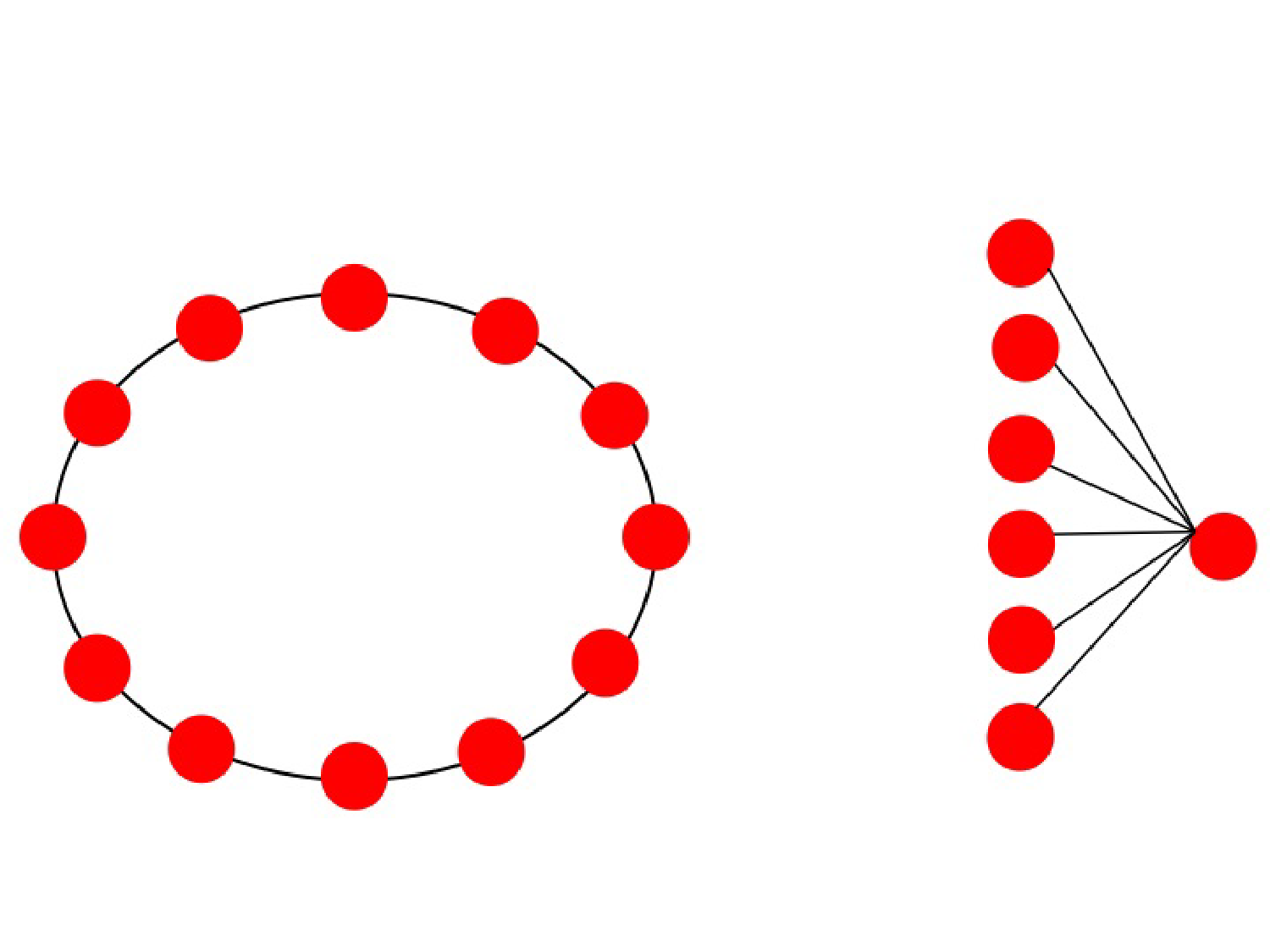,height=2.5in, width=2.5in}
\caption{An example illustrating the ``cycle \& star" lower bound construction for PageRank computations.}
\label{fig:pageranklower}
\end{center}
\end{figure}
\end{proof}

{\blue
\begin{theorem}[Graphs with Many Significant Nodes]\label{lowerboundpagerank}
Let $\alpha = 0.5$, $\Delta$ be integral and $c$ be given. Then, there are infinitely many $n$ such that there exists a graph on $n$ nodes where the output to {\sc SignificantPageRanks} on that graph has size $\Omega(\frac{n}{\Delta})$.
\end{theorem}

\begin{proof}
The construction is a variant of the one used in the proof of Theorem~\ref{lowerboundpagerank}. The graph is made of $\frac{n}{(3\Delta+1)}$ identical copies of an undirected star graph on $d+1 = 3\Delta$ nodes.
An easy calculation with the PageRank equations shows that each hub has PageRank of $\Delta + \frac{1}{3}$ and each leaf has PageRank of $\frac{2}{3}+\frac{1}{(9\Delta-3)} \leq 1$. The number of nodes with PageRank at least $\Delta$ is therefore $\frac{n}{d} = \Omega(\frac{n}{\Delta})$.

\end{proof}
}

\section{Acknowledgments}
{\blue
We would like to thank the anonymous reviewers of Internet Mathematics and WAW 2012 for their valuable feedback and comments. We thank Yevgeniy Vorobeychik, Elchanan Mossel and Brendan Lucier for their suggestions in early stages of this work.
}

\bibliographystyle{plain}
\bibliography{pagerank}

\appendix
\section{Concentration Bounds}
\begin{lemma}(\text{Multiplicative Chernoff Bound})
\label{lemma.multchernoff}
Let $X = \sum_{i=1}^{n}{X_i}$ be a sum of independent (but not necessarily identical) Bernoulli random variables. 
Then,
\begin{enumerate}
\item
For $0 < \lambda < 1$,
\begin{eqnarray*}
Pr[X < (1-\lambda)E[X]] &< \exp(-\frac{\lambda^2}2 E[X]) \\
Pr[X > (1+\lambda)E[X]] &< \exp(-\frac{\lambda^2}4 E[X]).
\end{eqnarray*}
\item
For $\lambda \geq 1$,
\[ \Pr[X > (1+\lambda)E[X]] < \exp(- \frac{\lambda E[X]}{3}).\]   
\item
For any constant $\Delta \geq (1+\lambda)E[X]$,
\begin{equation*} \label{eqn:Chernov}
Pr[X > \Delta] < \left\{
	\begin{array}{ll}
		\exp(-\frac{\lambda^2}4 \cdot \frac{\Delta}{1+\lambda})  & \mbox{if } 0< \lambda < 1 \\
		 \exp(- \frac{\lambda}{3} \cdot \frac{\Delta}{(1+\lambda)}) & \mbox{if } \lambda \geq 1
	\end{array}
\right.
\end{equation*}
\end{enumerate}
\end{lemma}

\begin{proof}
The case of $0 < \lambda <1$ is standard and a proof can be found, for example, in chapter 4 of \cite{MR05}.
For any $\lambda$, it is also shown therein that \[ \Pr[X > (1+\lambda)\mu n] \leq \left(\frac{e^{\lambda}}{{(1+\lambda)}^{(1+\lambda)}} \right)^{\mu n}.\] Now for $\lambda \geq 1$,  \[ \frac{e^{\lambda}}{{(1+\lambda)}^{(1+\lambda)}} < \exp \left(-\frac{\lambda^2}{2+\lambda} \right) \leq  \exp \left(-\frac{\lambda}{3} \right),\] and the second claimed item follows.

We now prove the last claimed item.
Assume that $\frac{\Delta}{(1+\lambda)} - E[X] > 0$ (otherwise the proof follows immediately from part 1). Define $k = \lceil{ \frac{\Delta}{(1+\lambda)} - E[X] \rceil}$ and $Y= \sum_{i=1}^{n+k} Y_i$, where for $1 \leq i \leq n$, $Y_i = X_i$  and for $ n < i \leq n+k $, $Y_i$ are independently distributed Bernoulli random variables with expectation
$(\frac{\Delta}{(1+\lambda)} - E[X])/k$ each. Note that $k \geq 1$, $Y_i$ are indeed Bernoulli random variables as $0 < (\frac{\Delta}{1+\lambda} - E[X])/k \leq 1$, and that $E[Y] = E[X]+ (\frac{\Delta}{(1+\lambda)} - E[X]) = \frac{\Delta}{(1+\lambda)}$.
Now,
\begin{align*}
& Pr \large (X > \Delta \large) = Pr \large(X > (1+\lambda)\Delta/(1+\lambda) \large) \leq \\
& Pr \large(Y > (1+\lambda)\Delta/(1+\lambda) \large)  < \left\{
	\begin{array}{ll}
		\exp(-\frac{\lambda^2}4 \cdot \frac{\Delta}{1+\lambda})  & \mbox{if } \lambda < 1 \\
		 \exp(- \frac{\lambda}{3} \cdot \frac{\Delta}{(1+\lambda)}) & \mbox{if } \lambda \geq 1
	\end{array}
\right.
\end{align*}

The next to last inequality follows from the fact that $Y$ first-order stochastically dominates $X$,
and the last inequality follows from parts 1 and 2 of the lemma.
\end{proof}

\end{document}